# General properties of the Solutions to Moving Boundary Problems for Black-Sholes Equations

Hyong-Chol O, Tae-Song Choe

Faculty of Mathematics, **Kim Il Sung** University, Pyongyang, D P R Korea

* Corresponding author's Email address: hc.o@ryongnamsan.edu.kp

**Abstract**: We study general properties such as the solution representation of a moving boundary value problem of the Black-Scholes equation, its min-max estimation, lower and upper gradient estimates, and strict monotonicity with respect to the spatial variables of the solution. These results are used in the study of a structural model of pricing puttable bond with credit risk. We first prove the solution representation of a special fixed boundary value problem of the Black-Scholes equation, the min-max estimate, the lower and upper gradient estimates, and the strict monotonicity with respect to the spatial variables of the solution. Then, these results are applied to give the solution representation of a moving boundary value problem of the Black-Scholes equation with moving boundary in the form of an exponential function, the min-max estimate, the lower and upper gradient estimates, and the strict monotonicity results on the spatial variables of the solution. Finally, we illustrate how these results can be used in the derivation of analytical pricing formulae and financial analysis of price functions of puttable bonds with credit risk (corporate bonds with one early redemption date). Our results can be used for the derivation and analysis of the analytical pricing formulae of the one-factor structural model of a more general puttable bonds with credit risk (corporate bond with several early redemption dates).

## 1. Introduction

Since the famous Black-Scholes formula in [ BS-1973 ], the Black-Scholes equation has become one of the most important tools in pricing options and many other financial derivatives. In this paper, we study the general properties of solutions such as solution representation of a moving boundary value problem of Black-Scholes partial differential equations, min-max estimates, gradient estimates with respect to spatial variables, and strict monotonicity of solutions.

It is well known that standard call and put options have pricing formulae (solution representations to the special terminal value problems of the Black-Scholes equation) and that the price function has various monotonicity (with respect to the stock price, the exercise price, the short-term interest rate, the dividend rate, and the volatility) and convexity (with respect to the stock price and the exercise price), and it has been studied that the price function of any European contingent claim as well as call options, inherits qualitative properties of maturity payoff functions (this means that the solution of the Black-Scholes equation inherits qualitative properties such as monotonicity or convexity of maturity payoff functions.[Jia-2005, Mer-1973, CR-1976, BGW-1996]. On the other hand, in [OK-2013 ], using the integral representation of solutions of the Black-Scholes equation with arbitrary maturity payoffs, a strict gradient estimate needed in the study of several expiry exotics was obtained. Recently, [OJK-2016] studied the general properties of solutions to the terminal value problem of the Black-Scholes equation with discontinuous maturity payoffs or inhomogeneous term, such as solution representation, min-max estimation, gradient estimation, strict monotonicity and convexity results with respect to the spatial variables of the solution, and using this strict monotonicity with respect to the spatial variables of the solution [OKP-2014, OJKJ-2017] obtained analytical pricing formulae for the 1 factor and 2 factor unified model of structural and reduced form approaches of corporate bond with discrete coupon.

On the other hand, pricing models of barrier options, the structural model or the unified model of structural and reduced form approaches for corporate bond prices with credit risk become the terminal-boundary value problems of the Black-Scholes equation [Jia-2005, OW-2013, Agl-2016]. Recently, in several literatures including [FNV-2016, RF-2016, RF-2017, ZW-2012, ZYL-2011 ], pricing bond options has been studied and the study on bond option with credit risk has been started. In [OKC2021], a simple structural model for pricing a puttable bond with a credit risk (a corporate bond with one early redemption date) is proposed and an analytical pricing formula for the bond is given, where solution representations of special boundary value problems of the Black-Scholes equations and a strict monotonicity result of the price function are used. Unlike the strict gradient estimates of the solution to the terminal value problem of the Black-Scholes equation of [OJK-2016 ], even the gradient estimate for the solution to a special boundary value problem for the Black-Scholes equation is not easily obtained from the integral representation of the solution, but is obtained from a rigorous analysis on the cumulative density function of the normal distribution and its derivatives, and the convexity of solution does not hold generally(see Theorem 3.2 bellow). In order to study the structural model for pricing corporate bonds with (several) early redemption dates and analyze the results, the results such as the solution representation of general boundary value problems of the Black-Scholes equation, the max-min estimate and a strict gradient estimate of the solution are needed, but it is difficult to find such results except for [Agl-2016].

The aim of this paper is to provide a max-min estimate and gradient estimate of the solution to the moving boundary value problem of the Black-Scholes equation arising in the study of pricing corporate bonds with early redemption provisions such as puttable bonds or callable bonds.

The rest of this paper is organized as follows. In Section 2, we give lemmas showing some properties of the cumulative density function of a one-dimensional normal distribution and its derivatives. In Section 3, we give a solution representation of a special fixed boundary value problem of the Black-Scholes equation, a strict max-min estimate, a lower and upper gradient estimate, and some sufficient conditions for the solution to increase strictly. Applying this result, Section 4 gives the solution representation of a moving boundary value problem of the Black-Scholes equation with moving boundary in the form of an exponential function, a strict max-min estimation, the lower and upper gradient estimates for the solution, and the conditions for the derivative to be strictly positive. Using the results of Sections 3 and 4, Section 5 introduces the 1 factor structural model for pricing puttable bond with credit risk (corporate bonds with one early redemption date)[OKC-2021], proves that the bond price function strictly increases, and financially analyzes the price formula given in [ OKC-2021 ] to derive the survival probability, the early redemption probability of and the default probability. Section 6 gives some conclusions.

## 2. Some Auxiliary Results

**Definition 2.1.** Denote the cumulative distribution function of one-dimensional normal distribution as $N(x) = \frac{1}{\sqrt{2\pi}} \int_{-\infty}^{x} e^{-\xi^2/2} d\xi$, we define the following two functions:

$$h(\xi) = (\sqrt{2\pi})^{-1} e^{-\xi^2/2} + \xi N(\xi), \xi < 0$$

$$A(\xi) := (\sqrt{2\pi})^{-1} e^{-\xi^2/2} / (-\xi N(\xi)), \xi < 0$$

**Lemma 2.1.** $h(\xi) = (\sqrt{2\pi})^{-1} e^{-\xi^2/2} + \xi N(\xi) > 0, \xi < 0.$

**Proof:** The required result is easily proved from

$$h(0) = (\sqrt{2\pi})^{-1} > 0, \ h(-\infty) = 0, \ h'(\xi) = N(\xi) > 0 \ .\text{(QED)}$$

**Lemma 2.2.** $A(-\infty) = 1$, $A(0-) = +\infty$, $A'(\xi) > 0 \ (\xi < 0)$ *and* $A : (-\infty, 0) \to (1, \infty)$ *is one to one correspondence.*

**Proof:** Using L'Hospital's rule, we have

$$\lim_{x \to -\infty} \frac{N(\xi)}{\xi e^{-\xi^2/2}} = \lim_{x \to -\infty} \frac{(\sqrt{2\pi})^{-1} e^{-\xi^2/2}}{e^{-\xi^2/2}(1 - \xi^2)} = 0,$$

and thus we have

$$\lim_{x \to -\infty} A(\xi) = \lim_{x \to -\infty} \frac{(\sqrt{2\pi})^{-1} e^{-\xi^2/2}}{-\xi N(\xi)} = -\lim_{x \to -\infty} \frac{(\sqrt{2\pi})^{-1} e^{-\xi^2/2}(-\xi)}{N(\xi) + \xi(\sqrt{2\pi})^{-1} e^{-\xi^2/2}} = 1.$$

On the other hand

$$A'(\xi) = \frac{e^{-\xi^2/2}[\xi^2 N(\xi) + N(\xi) + (\sqrt{2\pi})^{-1} \xi e^{-\xi^2/2}]}{\sqrt{2\pi} \xi^2 [N(\xi)]^2}$$

and here denote $z(\xi) := \xi^2 N(\xi) + N(\xi) + \xi e^{-\xi^2/2}$, then $z(0) = 1/2, z(-\infty) = 0$ and $z'(\xi) = 2h(\xi)$ and thus from Lemma 2.1, we have $z'(\xi) > 0$. So $A'(\xi) > 0$. (QED)

**Lemma 2.3.** *For any* $B > 1$, *denote* $\xi(B) = A^{-1}(B)$, *then we have the inequalities*

$$(\sqrt{2\pi})^{-1} e^{-\xi^2/2} + B\xi N(\xi) > 0, \ \xi(B) < \xi < 0$$

$$(\sqrt{2\pi})^{-1} e^{-\xi^2/2} + B\xi N(\xi) < 0, \ \xi < \xi(B) \ .$$

*(See Fig 1.)*

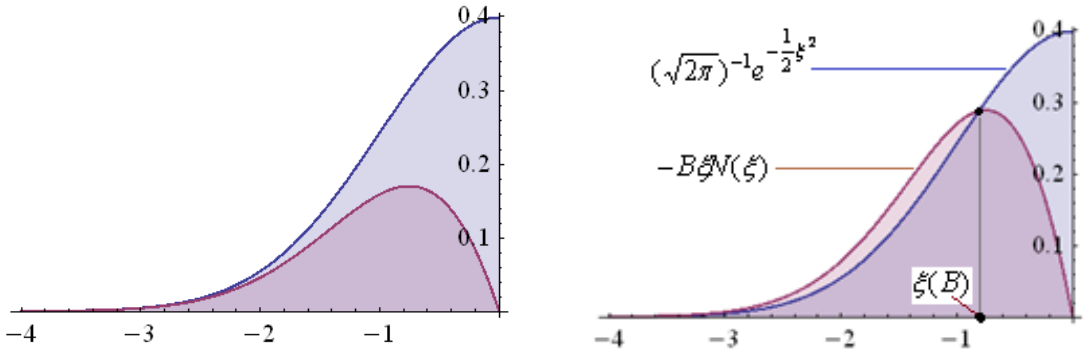

Fig. 1. The graphs of $(\sqrt{2\pi})^{-1} e^{-\xi^2/2}$ and $-B\xi N(\xi)$ (left: B=1, right: B=1.7)

**Proof:** From the definition of $\xi(B)$ and $A(\xi(B)) = B$, we have

$$(\sqrt{2\pi})^{-1} e^{-\xi^2(B)/2} + B\xi(B) N(\xi(B)) = 0.$$

And denote $h_B(\xi) := (\sqrt{2\pi})^{-1} e^{-\xi^2/2} + B\xi N(\xi)$, then $h'_B(\xi) = (B-1)(\sqrt{2\pi})^{-1} e^{-\xi^2/2} + BN(\xi) > 0$.

Thus we obtain the required result. (QED)

**Remark 2.1.** These lemmas show a very interesting property of the cumulative density function of normal distribution. $B = 1$ is the upper bound of the set of $B$ such that the inequality $(\sqrt{2\pi})^{-1} e^{-\xi^2/2} + B\xi N(\xi) > 0$ holds for all $\xi < 0$, and if $B > 1$, then this strict inequality holds only in the subinterval $\xi(B) < \xi < 0$.

## 3. Estimates on the solution to a special boundary value problem of the Black-Scholes equation

Consider the following special boundary value problem problem

$$\frac{\partial w}{\partial t} + \frac{1}{2}\sigma^2 x^2 \frac{\partial^2 w}{\partial x^2} + cx\frac{\partial w}{\partial x} = 0, \quad x > 1, 0 \leq t < T \tag{3.1}$$

$$w(1, t) = 0, \qquad\qquad 0 \leq t < T \tag{3.2}$$

$$w(x, T) = f(x), \qquad\qquad x > 1 \tag{3.3}$$

which is important in studying a moving boundary value problem of the Black-Scholes equation.

**Theorem 3.1.** (*Solution representation and min-max estimates*) *Assume that there are nonnegative constants $M, \alpha$ such that $|f(x)| \leq Mx^{\alpha \ln x}$. Then the solution to the terminal value problem*

$$\begin{cases} \dfrac{\partial u}{\partial t} + \dfrac{1}{2}\sigma^2 x^2 \dfrac{\partial^2 u}{\partial x^2} + cx\dfrac{\partial u}{\partial x} = 0, & x > 0, 0 \leq t < T \\ u(x, T) = f(x) \cdot 1\{x > 1\}, & x > 0 \end{cases}$$

*is represented as follows*

$$u(x, t) = \frac{1}{\sigma\sqrt{2\pi\tau}} \int_1^{+\infty} e^{-\frac{1}{2}d^2(x/z, t)} \frac{f(z)}{z} dz$$

$$\tau := T - t, \quad d(x, t) := (\sigma\sqrt{\tau})^{-1}[\ln x + (c - \sigma^2/2)\tau]. \tag{3.4}$$

*Then*

$$w(x, t) = u(x, t) - x^\beta u(1/x, t) \tag{3.5}$$

*is the solution of (3.1), (3.2), (3.3), where $\beta := 1 - 2c/\sigma^2$. Furthermore, if $A \leq f(x) \leq B$,*

then we have
$$A \cdot [N(d(x,t)) - x^\beta N(d(1/x,t))] \le w(x,t) \le B \cdot [N(d(x,t)) - x^\beta N(d(1/x,t))] \quad (3.6)$$
*and furthermore, if* $|\{x: f(x) > A\}| > 0$, *then we have*
$$A \cdot [N(d(x,t)) - x^\beta N(d(1/x,t))] < w(x,t) \quad (t < T)$$
*and if* $|\{x: f(x) < B\}| > 0$, *then we have*
$$w(x,t) < B \cdot [N(d(x,t)) - x^\beta N(d(1/x,t))] \quad (t < T).$$

**Proof**. (3.4) is (3) of Proposition 1 of [OK-2013], and (3.5) follows from the image solution method of [Bu-2001]. Then we have

$$w(x,t) = \frac{1}{\sigma\sqrt{2\pi\tau}} \int_1^{+\infty} \left( e^{-\frac{1}{2}[d(x/z,t)]^2} - x^\beta e^{-\frac{1}{2}[d(1/x/z,t)]^2} \right) \frac{f(z)}{z} dz. \quad (3.5)'$$

First we prove that $e^{-[d(x/z,t)]^2/2} - x^\beta e^{-[d(1/x/z,t)]^2/2} > 0 \, (x>1, z>1)$. Considering

$$d(x/z) = d(x) - \frac{\ln z}{\sigma\sqrt{\tau}} \quad \text{and} \quad e^{-\frac{1}{2}[d(x) - \frac{\ln z}{\sigma\sqrt{\tau}}]^2} = e^{-\frac{1}{2}[d(1/x) - \frac{\ln z}{\sigma\sqrt{\tau}}]^2} e^{-\frac{1}{2}\{[d(x) - \frac{\ln z}{\sigma\sqrt{\tau}}]^2 - [d(1/x) - \frac{\ln z}{\sigma\sqrt{\tau}}]^2\}},$$

we have $e^{-\frac{1}{2}[d(x) - \frac{\ln z}{\sigma\sqrt{\tau}}]^2} = x^{\beta + 2\ln z/(\sigma^2\tau)} e^{-\frac{1}{2}[d(1/x) - \frac{\ln z}{\sigma\sqrt{\tau}}]^2}$ and thus we have

$$e^{-\frac{1}{2}[d(x) - \frac{\ln z}{\sigma\sqrt{\tau}}]^2} - x^\beta e^{-\frac{1}{2}[d(1/x) - \frac{\ln z}{\sigma\sqrt{\tau}}]^2} = e^{-\frac{1}{2}[d(1/x) - \frac{\ln z}{\sigma\sqrt{\tau}}]^2} x^\beta (x^{2\ln z/(\sigma^2\tau)} - 1) > 0 \, (x>1, z>1).$$

Next considering $A \le f(x) \le B$ and $\frac{1}{\sigma\sqrt{2\pi\tau}} \int_1^{+\infty} e^{-\frac{1}{2}[d(x,t) - \frac{\ln z}{\sigma\sqrt{\tau}}]^2} \frac{dz}{z} = N(d(x,t))$, then we have (3.6). (QED)

**Theorem 3.2**. *If* $f(x) \equiv 1$ *then the solution to the problem (3.1), (3.2), (3.3) is provided as follows:*
$$w(x,t) = N(d(x,t)) - x^\beta N(d(1/x,t)). \quad (3.7)$$
*Furthermore, we have the following estimates:*
$$0 = w(1,t) < w(x,t) < w(+\infty,t) = 1, \quad w_x(+\infty,t) = 0 < w_x(x,t) \le C \, (x \ge 1, \, t < T).$$
*In particular, if* $c \ge 0$, *then* $w_{xx}(x,t) < 0 \, (x > 1, \, t < T)$ *and thus* $w_x(x,t) < w_x(1,t)$. *If* $c < 0$, *then* $w_{xx}(x,t) < 0 \, (x > e^{-2c(T-t)}, \, t < T)$ *and thus* $w_x$ *has a positive maximum in the interval* $[1, e^{-2c(T-t)}]$.

**Proof**: (3.7) follows directly from (3.4) and (3.5), and $w(1,t) = 0, w(-\infty,t) = 1$ follows directly from (3.7). Considering $d_x(x,t) = (x\sigma\sqrt{\tau})^{-1}$, then we have

$$w_x(x,t) = \frac{e^{-d^2(x,t)/2}}{\sqrt{2\pi} \, x\sigma\sqrt{\tau}} - \beta x^{\beta-1} N(d(1/x,t)) + x^\beta \frac{e^{-d^2(1/x,t)/2}}{\sqrt{2\pi} \, x\sigma\sqrt{\tau}}.$$

Here considering
$$e^{-d^2(x,t)/2} = x^\beta e^{-d^2(1/x,t)/2}, \quad d(x,t) = (\sigma\sqrt{\tau})^{-1}\ln x - (\beta\sigma/2)\sqrt{\tau}, \tag{3.8}$$

We have
$$w_x(x,t) = \frac{2x^{\beta-1}}{\sigma\sqrt{\tau}}\left[\frac{e^{-d^2(1/x,t)/2}}{\sqrt{2\pi}} - \frac{\beta\sigma\sqrt{\tau}}{2} N\left(d\left(\frac{1}{x},t\right)\right)\right] \tag{3.9}$$

$$= \frac{2x^{\beta-1}}{\sigma\sqrt{\tau}}\left[\frac{e^{-d^2(1/x,t)/2}}{\sqrt{2\pi}} + d\left(\frac{1}{x},t\right)N\left(d\left(\frac{1}{x},t\right)\right) + \frac{\ln x}{\sigma\sqrt{\tau}}N\left(d\left(\frac{1}{x},t\right)\right)\right].$$

Here we have considered that from (3.8) we have $d(1/x,t) + (\sigma\sqrt{\tau})^{-1}\ln x = -\beta\sigma\sqrt{\tau}/2$). From this expression, $w_x(+\infty, t) = 0$ and the upper boundedness of the derivative follows directly. Also, from Lemma 2.1, the sum of the first two terms of the above expression is always positive and the last term is not negative for $x \geq 1$ and thus we have $w_x(x,t) > 0$ ($x \geq 1$). Hence we have $w(1,t) < w(x,t) < w(+\infty, t)$.

Differentiate (3.9) with respect to $x$, we have
$$w_{xx} = \frac{2x^{\beta-2}}{\sigma\sqrt{\tau}}\left[\frac{e^{-d^2(1/x,t)/2}}{\sqrt{2\pi}}\left(\beta - 1 + \frac{-\ln x}{\sigma^2\tau} - \frac{\beta\sigma\sqrt{\tau}/2}{\sigma\sqrt{\tau}} + \frac{\beta}{2}\right) - \frac{(\beta-1)\beta\sigma\sqrt{\tau}}{2}N\left(d\left(\frac{1}{x},t\right)\right)\right]$$

$$= \frac{-2x^{\beta-2}}{\sigma^3\sqrt{\tau}}\left\{\left(2c + \frac{\ln x}{\tau}\right)\frac{e^{-d^2(1/x,t)/2}}{\sqrt{2\pi}} - \frac{2c\beta\sigma\sqrt{\tau}}{2}N\left(d\left(\frac{1}{x},t\right)\right)\right\} \tag{3.10}$$

If $c \geq 0$, then $w_{xx}$ can be rewritten as
$$\frac{-2x^{\beta-2}}{\sigma^3\sqrt{\tau}}\left\{c\left[\frac{e^{-d^2(1/x,t)/2}}{\sqrt{2\pi}} + d\left(\frac{1}{x},t\right)N\left(d\left(\frac{1}{x},t\right)\right) + \frac{\ln x}{\sigma\sqrt{\tau}}N\left(d\left(\frac{1}{x},t\right)\right)\right] + \frac{\ln x}{\tau}\frac{e^{-d^2(1/x,t)/2}}{\sqrt{2\pi}}\right\}$$

and considering Lemma 2.1, then we have $w_{xx}(x,t) < 0$ for $x > 1$. If $c < 0$, then $\beta > 1$ and thus $-c\beta > 0$, and from (3.10) we have $w_{xx}(x,t) < 0$ when $x \geq e^{-2c\tau}$. Since $w_{xx}(1,t) > 0$, then in the interval $[1, e^{-2c(T-t)}]$, we have a unique inflection point (a point such that $w_{xx} = 0$)(see Fig. 2). (QED)

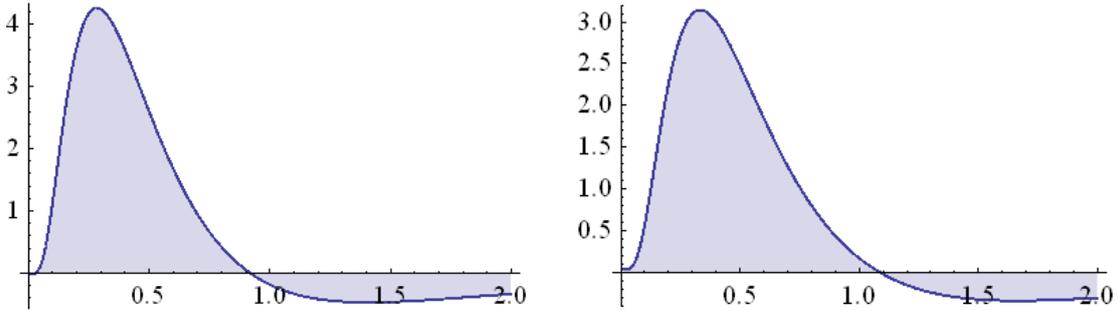

Fig. 2.　The graphs of $w_{xx}$ (left: the case that $c \geq 0$, right: the case that $c < 0$)

**Remark 3.1.** As shown in Theorem 3.2, the gradient estimates of the solution of the boundary value problem of the Black-Scholes equation , unlike the rigorous gradient estimates of the solution of the terminal value problem [OJK-2016, Theorem 2], are not easily obtained from the integral representation of the solution even in the case of special terminal values but from a rigorous analysis of the cumulative density function and its derivative of the normal distribution. Also, in the terminal value problem of the Black-Scholes equation, the convexity of the terminal value function is inherited to the solution [OJK-2016, Theorem 3], but in the boundary value problem, the solution may not be convex, even if the end value function is convex.

**Theorem 3.3** *(Gradient Representation) If $f(x)$ is continuous, piecewise differentiable, and there are nonnegative constants $M, \alpha$ such that $|f(x)| \leq M \cdot x^{\alpha \ln x}$, $|f'(x)| \leq M \cdot x^{\alpha \ln x}$, then the derivative $w_x$ of the solution $w$ of (3.1), (3.2), (3.3) can be represented as follows:*

$$w_x(x,t) = \frac{x^{\beta-1}}{\sqrt{2\pi}} \left\{ \frac{2f(1)}{\sigma\sqrt{\tau}} e^{-\frac{1}{2}d^2(1/x)} - \beta \int_{-\infty}^{d(1/x)} e^{-\frac{1}{2}\xi^2} f(e^{[d(1/x)-\xi]\sigma\sqrt{\tau}}) d\xi \right\} +$$

$$+ \frac{e^{c\tau}}{\sqrt{2\pi}} \left\{ \int_{-\infty}^{b(x)} e^{-\frac{1}{2}\xi^2} f'(e^{[b(x)-\xi]\sigma\sqrt{\tau}}) d\xi + x^{\beta-2} \int_{-\infty}^{b(1/x)} e^{-\frac{1}{2}\xi^2} f'(e^{[b(1/x)-\xi]\sigma\sqrt{\tau}}) d\xi \right\}$$

$$b(x) := d(x) + \sigma\sqrt{\tau} = (\sigma\sqrt{\tau})^{-1}[\ln x + (c + \sigma^2/2)\tau]. \tag{3.11}$$

**Proof**: Let $\tau = T - t$ in Eq. (3.4) and change the integral variable $z$ by
$$\xi = (\sigma\sqrt{\tau})^{-1}[\ln x - \ln z + (c - \sigma^2/2)\tau], \quad d(x) = (\sigma\sqrt{\tau})^{-1}[\ln x + (c - \sigma^2/2)\tau],$$
then we have $u(x,t) = (\sqrt{2\pi})^{-1} \int_{-\infty}^{d(x,t)} e^{-\xi^2/2} f(\exp[(d(x,t)-\xi)\sigma\sqrt{\tau}]) d\xi$ and (3.5)' are written as follows:

$$w(x,t) = \frac{1}{\sqrt{2\pi}} \left[ \int_{-\infty}^{d(x,t)} e^{-\frac{1}{2}\xi^2} f(e^{[d(x,t)-\xi]\sigma\sqrt{\tau}}) d\xi - x^\beta \int_{-\infty}^{d(1/x,t)} e^{-\frac{1}{2}\xi^2} f(e^{[d(1/x,t)-\xi]\sigma\sqrt{\tau}}) d\xi \right].$$

Consider $d'(x) = (x\sigma\sqrt{\tau})^{-1}$ and $[d(1/x)]' = -(x\sigma\sqrt{\tau})^{-1}$, and differentiate this expression then we have

$$w_x(x,t) = \frac{1}{x\sigma\sqrt{2\pi\tau}} \left[ e^{-d^2(x)/2} f(1) + x^\beta e^{-d^2(1/x)/2} f(1) \right] +$$

$$+ \frac{1}{x\sqrt{2\pi}} \int_{-\infty}^{d(x)} e^{-\xi^2/2} f'(e^{[d(x)-\xi]\sigma\sqrt{\tau}}) e^{[d(x)-\xi]\sigma\sqrt{\tau}} d\xi -$$

$$- \frac{1}{\sqrt{2\pi}} \beta x^{\beta-1} \int_{-\infty}^{d(1/x)} e^{-\xi^2/2} f(e^{[d(1/x)-\xi]\sigma\sqrt{\tau}}) d\xi +$$

$$+ \frac{1}{\sqrt{2\pi}} x^{\beta-1} \int_{-\infty}^{d(1/x)} e^{-\xi^2/2} f'(e^{[d(1/x)-\xi]\sigma\sqrt{\tau}}) e^{[d(1/x)-\xi]\sigma\sqrt{\tau}} d\xi.$$

Here consider $e^{-d^2(x)/2} = e^{-d^2(1/x)/2} x^\beta$ of (3.8) and $e^{-\xi^2/2} e^{[d(x)-\xi]\sigma\sqrt{\tau}} = e^{c\tau} x e^{-(\xi+\sigma\sqrt{\tau})^2/2}$, and denote by $b(x) = d(x) + \sigma\sqrt{\tau}$, then we have the gradient representation (3.11). (QED)

**Theorem 3.4.** *(Lower and Upper Estimates of Gradient) If $f(x)$ is continuous,*

*piecewise differentiable, and if* $B_1 \leq f(x) \leq B_2$ *and* $D_1 \leq f'(x) \leq D_2$, *then we have the lower and upper estimates on the gradient of the solution* $w$ *of* (3.1), (3.2), (3.3):

$$E_1(x,t) \leq w_x(x,t) \leq E_2(x,t). \tag{3.12}$$

*Here*

$$E_1(x,t) = x^{\beta-1}\left\{\frac{2e^{-d^2(1/x)/2}}{\sigma\sqrt{2\pi\tau}}f(1) + N\left[d\left(\frac{1}{x}\right)\right]\min(-\beta \cdot f)\right\} + e^{c\tau}\left\{N(b(x)) + x^{\beta-2}N\left[b\left(\frac{1}{x}\right)\right]\right\}\min f'$$

$$E_2(x,t) = x^{\beta-1}\left\{\frac{2e^{-d^2(1/x)/2}}{\sigma\sqrt{2\pi\tau}}f(1) + N\left[d\left(\frac{1}{x}\right)\right]\max(-\beta \cdot f)\right\} + e^{c\tau}\left\{N(b(x)) + x^{\beta-2}N\left[b\left(\frac{1}{x}\right)\right]\right\}\max f'$$

(3.13)

Proof: If $\beta \leq 0$, then the coefficient of the second term in (3.11) is not negative, and therefore denote as

$$E_1(x,t) = x^{\beta-1}\left[\frac{2f(1)}{\sigma\sqrt{2\pi\tau}}e^{-d^2(1/x)/2} - \beta B_1 N(d(1/x))\right] + e^{c\tau}D_1\left[N(b(x)) + x^{\beta-2}N(b(1/x))\right]$$

$$E_2(x,t) = x^{\beta-1}\left[\frac{2f(1)}{\sigma\sqrt{2\pi\tau}}e^{-d^2(1/x)/2} - \beta B_2 N(d(1/x))\right] + e^{c\tau}D_2\left[N(b(x)) + x^{\beta-2}N(b(1/x))\right]$$

and if $\beta > 0$, the coefficient of the second term in (3.11) is negative, and therefore denote as

$$E_1(x,t) = x^{\beta-1}\left[\frac{2f(1)}{\sigma\sqrt{2\pi\tau}}e^{-d^2(1/x)/2} - \beta B_2 N(d(1/x))\right] + e^{c\tau}D_1\left[N(b(x)) + x^{\beta-2}N(b(1/x))\right]$$

$$E_2(x,t) = x^{\beta-1}\left[\frac{2f(1)}{\sigma\sqrt{2\pi\tau}}e^{-d^2(1/x)/2} - \beta B_1 N(d(1/x))\right] + e^{c\tau}D_2\left[N(b(x)) + x^{\beta-2}N(b(1/x))\right],$$

Then we have (3.12). Combining the two cases yields (3.13).(QED)

**Remark 3.2.** The gradient representation (3.11) is very useful. From this, we can obtain a condition for the solution to be strictly increasing. If $\beta \leq 0$, we can easily obtain a condition for the solution to be strictly monotonically increasing from the lower and upper estimates (3.12) and (3.13), but if $\beta > 0$, then it does not happen easily.

**Corollary 3.5**. (*Strict monotonicity*) *Under the assumption of Theorem 3.3, assume that* $f(1) > 0, f'(x) \geq 0$. *Then if* $\beta \leq 0$, *we have always* $w_x > 0$ $(x > 1)$ *for the solution* $w$ *of* (3.1), (3.2), (3.3), *and if* $\beta > 0$, *then* $w_x > 0$ $(x > 1)$ *when the following holds.*

$$\frac{2f(1)}{\sigma\sqrt{\tau}}e^{-\frac{1}{2}d^2(\frac{1}{x})} + \frac{e^{c\tau}}{x^{\beta-1}}\left\{\int_{-\infty}^{b(x)}e^{-\frac{1}{2}\xi^2}f'(e^{[b(x)-\xi]\sigma\sqrt{\tau}})d\xi + x^{\beta-2}\int_{-\infty}^{b(1/x)}e^{-\frac{1}{2}\xi^2}f'(e^{[b(\frac{1}{x})-\xi]\sigma\sqrt{\tau}})d\xi\right\}$$

$$> \beta \int_{-\infty}^{d(1/x)}e^{-\frac{1}{2}\xi^2}f(e^{[d(\frac{1}{x})-\xi]\sigma\sqrt{\tau}})d\xi, x > 1, 0 < \tau \leq T$$

(3.14)

**Proof**: Since $f'(x) \geq 0$, the last two terms of (3.11) are nonnegative, so if $\beta \leq 0$ then we always have $w_x(x,t) > 0$. Now suppose that $\beta > 0$. Then from (3.11), $w_x(x,t) > 0$ if and only if the inequality (3.14) holds. (QED)

**Remark 3.3.** Although Corollary 3.4 gives a necessary and sufficient condition for $w_x > 0$ ($x > 1$) but when $\beta > 0$, it seems to be difficult to judge if (3.14) holds or not. Therefore, we consider a convenient method to judge.

Assuming that $f, f'$ are all bounded, we try to find a sufficient condition for the sum of the first two terms of (3.11) to be always positive. Denote the sum of the first two terms of (3.11) as $(\sqrt{2\pi})^{-1} x^{\beta-1} g(x)$. Then

$$g(x) = \frac{2f(1)}{\sigma\sqrt{\tau}} e^{-d^2(1/x)/2} - \beta \int_{-\infty}^{d(1/x)} e^{-\xi^2/2} f(e^{[d(1/x)-\xi]\sigma\sqrt{\tau}}) d\xi. \tag{3.15}$$

Here considering $d(1/x) = (\sigma\sqrt{\tau})^{-1}[(c-\sigma^2/2)\tau - \ln x]$, then we have

$$g(0+) = -\beta f(+\infty)\sqrt{2\pi} < 0, \quad g(+\infty) = 0. \tag{3.16}$$

From the assumption, $f(+\infty) > 0$ and thus $g(0+) < 0$.

First find the condition for $g(1) > 0$. From (3.15), $g(1) > 0$ if and only if

$$f(1) > \frac{\beta \cdot e^{d^2(1,\tau)/2} \sigma\sqrt{\tau}}{2} \int_{-\infty}^{d(1,\tau)} e^{-\xi^2/2} f(e^{[d(1,\tau)-\xi]\sigma\sqrt{\tau}}) d\xi (0 < \forall \tau \leq T).$$

Let us find a sufficient condition that does not depend on $\tau$. From the definitions of (3.4) and $\beta$, we have $d(1,\tau) = -\beta\sigma\sqrt{\tau}/2$. On the right-hand side of the above inequality, change the integral variable $\xi$ into $\eta = [-\beta \cdot \sigma\sqrt{\tau}/2 - \xi]\sigma\sqrt{\tau}$, then we have

$$\frac{\beta \cdot e^{\beta^2\sigma^2\tau/8}}{2} \sigma\sqrt{\tau} \int_{-\infty}^{-\beta\cdot\sigma\sqrt{\tau}/2} e^{-\xi^2/2} f(e^{[-\beta\cdot\sigma\sqrt{\tau}/2-\xi]\sigma\sqrt{\tau}}) d\xi =$$

$$= \frac{\beta \cdot e^{\beta^2\sigma^2\tau/8}}{2} \int_0^\infty e^{-[\beta\cdot\sigma\sqrt{\tau}/2+\eta/\sigma\sqrt{\tau}]^2/2} f(e^\eta) d\eta = \frac{\beta}{2} \int_0^\infty e^{-\frac{\eta^2}{2\sigma^2\tau}} e^{-\beta\eta/2} f(e^\eta) d\eta.$$

This expression increases on $\tau$. Thus we obtained a sufficient condition for $g(1) > 0$:

$$f(1) > \frac{\beta\sigma\sqrt{T} e^{\beta^2\sigma^2 T/8}}{2} \int_{-\infty}^{-\beta\cdot\sigma\sqrt{T}/2} e^{-\xi^2/2} f(e^{[-\beta\cdot\sigma\sqrt{T}/2-\xi]\sigma\sqrt{T}}) d\xi = \frac{\beta}{2} \int_0^\infty e^{-\frac{\eta^2}{2\sigma^2 T} - \frac{\beta\eta}{2}} f(e^\eta) d\eta. \tag{3.17}$$

To find a more convenient criteria, consider the following inequality

$$\frac{\beta\sigma\sqrt{T}e^{\beta^2\sigma^2 T/8}}{2}\int_{-\infty}^{-\beta\cdot\sigma\sqrt{T}/2}e^{-\xi^2/2}f(e^{[-\beta\cdot\sigma\sqrt{T}/2-\xi]\sigma\sqrt{T}})d\xi \leq f(+\infty)\frac{\beta\sigma\sqrt{T}e^{\beta^2\sigma^2 T/8}}{2}\int_{-\infty}^{-\beta\cdot\sigma\sqrt{T}/2}e^{-\xi^2/2}d\xi$$

$$= f(+\infty)\frac{\beta\sigma\sqrt{T}/2\cdot N(-\beta\sigma\sqrt{T}/2)}{(\sqrt{2\pi})^{-1}e^{-(\beta\sigma\sqrt{T}/2)^2/2}} = f(+\infty)[A(-\beta\sigma\sqrt{T}/2)]^{-1}.$$

Here $A(\xi)(>1)$ is the function defined in Section 2. Then from
$$f(+\infty) \leq A(-\beta\sigma\sqrt{T}/2)\cdot f(1), \tag{3.18}$$
we directly obtain (3.17).

Now we will prove that there is a unique $x_0 \geq 1$ such that $g'(x_0) = 0$ and if $1 \leq x < x_0$ then $g'(x) > 0$, and if $x > x_0$ then $g'(x) < 0$. From this result, we have $g(x_0) = \max_{[1,+\infty)} g(x) > 0$, and hence we have $g(x) > 0, x \geq 1$ from the last equality of (3.16), and from this we have $w_x(x,t) > 0 (x > 1)$.

Calculating directly from (3.15), we obtain the following
$$g'(x) = \frac{1}{x}\left\{ e^{-\frac{1}{2}d^2(1/x)}\left[\frac{2f(1)}{\sigma^2\tau}\frac{-\ln x + (c-\sigma^2/2)\tau}{\sigma\sqrt{\tau}} + \beta\frac{f(1)}{\sigma\sqrt{\tau}}\right] \right.$$
$$\left. + \beta\int_{-\infty}^{d(1/x)} e^{-\frac{\xi^2}{2}} f'(e^{[d(1/x)-\xi]\sigma\sqrt{\tau}})e^{[d(1/x)-\xi]\sigma\sqrt{\tau}} d\xi \right\}.$$

Hence
$$g'(x) = \frac{1}{x}\left\{ \frac{-2f(1)}{(\sigma\sqrt{\tau})^3}\ln x \cdot e^{-\frac{1}{2}d^2(1/x)} + \beta\int_{-\infty}^{d(1/x)} e^{-\frac{\xi^2}{2}} f'(e^{[d(1/x)-\xi]\sigma\sqrt{\tau}})e^{[d(1/x)-\xi]\sigma\sqrt{\tau}} d\xi \right\}, x > 0.$$

Obviously $g'(+\infty) = 0$ and $g'(x) > 0$ for $0 < x < 1$. In particular, if $|\{x: f'(x) > 0\}| > 0$, then $g'(1) > 0$. Define the following functions:
$$\underline{g}'(x) = \frac{1}{x}\cdot\frac{-2f(1)}{(\sigma\sqrt{\tau})^3}\ln x e^{-\frac{1}{2}d^2(1/x)},$$

$$\overline{g}'(x) = \frac{1}{x}\left\{ \frac{-2f(1)}{(\sigma\sqrt{\tau})^3}\ln x e^{-\frac{1}{2}d^2(1/x)} + \beta D\int_{-\infty}^{d(1/x)} e^{-\frac{\xi^2}{2}} e^{[d(1/x)-\xi]\sigma\sqrt{\tau}} d\xi \right\} (D = \max f')$$

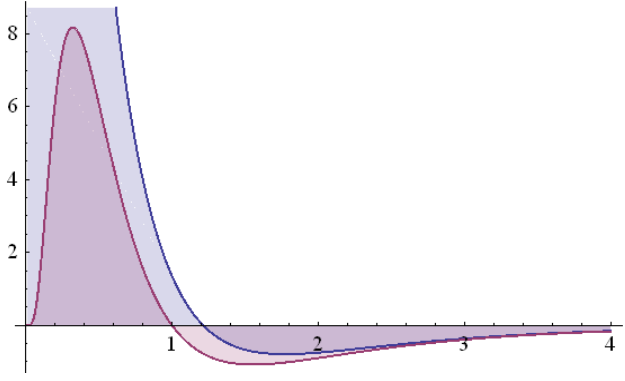

그림3. $\underline{g}'(x)$ 와 $\overline{g}'(x)$ 의 그래프 ($c = 0.03, \sigma = 0.5, \tau = 2, D/f(1) = 1$)

Then $\underline{g}'(x) < g'(x) < \overline{g}'(x)$ and if $0 < x < 1$ then $\underline{g}'(x) > 0$ and $\underline{g}'(1) = 0$, and if $x > 1$ then $\underline{g}'(x) < 0$. Since $-\xi^2/2 - \xi\sigma\sqrt{\tau} = -(\xi + \sigma\sqrt{\tau})^2/2 + \sigma^2\tau/2$, we have

$$\int_{-\infty}^{d(1/x)} e^{-\frac{\xi^2}{2}} e^{[d(\frac{1}{x})-\xi]\sigma\sqrt{\tau}} d\xi = e^{d(\frac{1}{x})\sigma\sqrt{\tau}+\frac{\sigma^2\tau}{2}} \int_{-\infty}^{d(1/x)} e^{-\frac{1}{2}(\xi+\sigma\sqrt{\tau})^2} d\xi = \frac{e^{c\tau}}{x}\sqrt{2\pi} N\left(d\left(\frac{1}{x}\right)+\sigma\sqrt{\tau}\right)$$

$$e^{-\frac{1}{2}d^2(1/x)} = e^{-\frac{1}{2}\left[d\left(\frac{1}{x}\right)+\sigma\sqrt{\tau}\right]^2} e^{\frac{1}{2}\left\{\left[d\left(\frac{1}{x}\right)+\sigma\sqrt{\tau}\right]^2 - d^2\left(\frac{1}{x}\right)\right\}} = e^{-\frac{1}{2}\left(d\left(\frac{1}{x}\right)+\sigma\sqrt{\tau}\right)^2} x^{-1} e^{c\tau}.$$

And using the notation $b(x) := d(x) + \sigma\sqrt{\tau} = (\sigma\sqrt{\tau})^{-1}[\ln x + (c+\sigma^2/2)\tau]$ of (3.11), $\overline{g}'(x)$ can be rewritten as follows

$$\overline{g}'(x) = \frac{e^{c\tau}}{x^2}\left\{\frac{-2f(1)}{(\sigma\sqrt{\tau})^3}\ln x \cdot e^{-b^2(1/x)/2} + \beta D \cdot \sqrt{2\pi} N(b(1/x))\right\}.$$

Thus $\overline{g}'(x) > 0$ ($0 < x \le 1$), and $\overline{g}'(x) \le 0$ is possible when $x > 1$. For $x > 1$, $\overline{g}'(x)$ can be rewritten as follows

$$\overline{g}'(x) = \frac{2\sqrt{2\pi}f(1)e^{c\tau}}{(\sigma\sqrt{\tau})^3} \frac{-\ln x}{x^2}\left\{\frac{e^{-b^2(1/x)/2}}{\sqrt{2\pi}} + \frac{\beta D}{2f(1)}\frac{(\sigma\sqrt{\tau})^3}{-\ln x} N(b(1/x))\right\}.$$

Then since $-\ln x < 0$, we have

$$\overline{g}'(x) > 0 \Leftrightarrow \ln x < \frac{\sqrt{2\pi}\beta D(\sigma\sqrt{\tau})^3}{2f(1)} \frac{N(b(1/x))}{e^{-b^2(1/x)/2}}.$$

Considering $\frac{N(b(1/x))}{e^{-b^2(1/x)/2}} \downarrow 0$ as $x \to +\infty$, there exists a $x_1 > 1$ such that $\overline{g}'(x_1) = 0$ and $\overline{g}'(x) > 0$ when $1 < x < x_1$, and $\overline{g}'(x) < 0$ when $x > x_1$ (see Fig. 3). Considering

$$\overline{g}'(x) - \underline{g}'(x) = \frac{1}{x}\beta D \int_{-\infty}^{d(1/x)} e^{-\xi^2/2} e^{[d(1/x)-\xi]\sigma\sqrt{\tau}} d\xi \downarrow 0,\ x \uparrow +\infty,$$

there exists a unique $x_0 (1 \le x_0 < x_1)$ such that $g'(x_0) = 0$ and $g'(x) > 0$ when $1 \le x < x_0$, and $g'(x) < 0$ when $x > x_0$. Thus we proved the following theorem:

**Theorem 3.6** (*Strict monotonicity*) *Assume that $f(x)$ is continuous, piecewise differentiable, and that there are nonnegative constants $M, \alpha$ such that $0 \le f'(x) \le D$, $0 < f(1) \le f(x) \le f(+\infty) < +\infty$, $|\{x: f'(x) > 0\}| > 0$, $\beta > 0$. Then we have $w_x > 0\ (x > 1)$ if the inequality* (3.17) *holds. Especially if the inequality* (3.18) *holds, then* (3.17) *holds.*

## 4. Estimates of the solution to a moving boundary value problem of the Black-Scholes equation

Using the results of the previous section, we can obtain the max-min estimates of the solutions to the moving boundary value problem of the general Black-Scholes equation.

$$\frac{\partial V}{\partial t} + \frac{1}{2}\sigma^2 S^2 \frac{\partial^2 V}{\partial S^2} + (r-q)S\frac{\partial V}{\partial S} - rV = 0,\ S > be^{-a(T-t)}, 0 \le t < T \quad (4.1)$$

$$V(be^{-a(T-t)}, t) = 0, \qquad\qquad 0 \le t < T \quad (4.2)$$

$$V(S, T) = g(S), \qquad\qquad S > be^{-a(T-t)}. \quad (4.3)$$

Now using the transformation of

$$x = S/[be^{-a(T-t)}],\ w(x,t) = V(S,t)/e^{-r(T-t)},$$

we have

$$V_t = e^{-r(T-t)}(rw + w_t - axw_x),\ SV_S = e^{-r(T-t)}xw_x,\ S^2 V_{SS} = e^{-r(T-t)}x^2 w_{xx}$$

and thus we have

$$\frac{\partial w}{\partial t} + \frac{1}{2}\sigma^2 x^2 \frac{\partial^2 w}{\partial x^2} + (r-q-a)x\frac{\partial w}{\partial x} = 0,\ x > 1, 0 \le t < T \quad (4.4)$$

$$w(1, t) = 0, \qquad\qquad 0 \le t < T \quad (4.5)$$

$$w(x, T) = g(bx) =: f(x), \qquad x > 1. \quad (4.6)$$

This is just the problem (3.1)~(3.3) when $c = r - q - a$, so $w(x,t)$ is given by (3.5) when $\beta := 1 - 2(r-q-a)/\sigma^2$. Thus, from Theorem 3.1, we obtain the solution representation and the min-max estimate of (4.1) to (4.3). To simplify the notation below, we write $x = S/[be^{-a(T-t)}]$.

**Theorem 4.1.** (*Solution representation and min-max estimates*) *Assume that there exist nonnegative constants $M, \alpha$ such that $|g(x)| \le Mx^{\alpha \ln x}$. Then the solution $V$ to the problem* (4.1), (4.2), (4.3) *is represented as follows.*

$$V(S,t) = \frac{e^{-r\tau}}{\sigma\sqrt{2\pi\tau}} \int_1^{+\infty} \left\{ e^{-\frac{1}{2}d^2(x/z)} - x^\beta e^{-\frac{1}{2}d^2(x^{-1}/z)} \right\} \frac{g(bz)}{z} dz$$

$$= \frac{e^{-r\tau}}{\sqrt{2\pi}} \left[ \int_{-\infty}^{d(x)} e^{-\frac{\xi^2}{2}} g(be^{[d(x)-\xi]\sigma\sqrt{\tau}}) d\xi - x^\beta \int_{-\infty}^{d(1/x)} e^{-\frac{\xi^2}{2}} g(be^{[d(1/x)-\xi]\sigma\sqrt{\tau}}) d\xi \right]$$

$$\tau = T-t, \quad d(x,t) := (\sigma\sqrt{\tau})^{-1}[\ln x + (r-q-a-\sigma^2/2)\tau]. \tag{4.7}$$

*Furthermore, if $g(S)$ is bounded, then we have the following min-max estimates*:

$$\min g \cdot e^{-r\tau}[N(d(x,t)) - x^\beta N(d(1/x))] \leq V(S,t) \leq \max g \cdot e^{-r\tau}[N(d(x,t)) - x^\beta N(d(1/x))]. \tag{4.8}$$

*In particular, if $|\{x : g(x) > \min g\}| > 0$ or $|\{x : g(x) < \max g\}| > 0$, then the strict inequalities in (4.8) hold.*

**Remark 4.1.** In [Agl-2016] they obtained the integral representation of solution and min-max estimates.

From Theorem 3.3 and $V_S = b^{-1}e^{-(r-a)(T-t)}w_x$, $f(x) = g(bx)$, $f'(x) = bg'(bx)$, we have the following gradient representation and the lower and upper estimates.

**Theorem 4.2.** (*Gradient representation*) *If $g(S)$ is continuous, piecewise differentiable, and there are nonnegative constants $M, \alpha$ such that $|g(S)| \leq M \cdot S^{\alpha \ln S}$, $|g'(S)| \leq M \cdot S^{\alpha \ln S}$, then the derivative $V_S$ of the solutions $V$ of (4.1), (4.2), (4.3) is represented as follows.*

$$V_S(S,t) = \frac{e^{-r\tau}}{\sqrt{2\pi}be^{-a\tau}} x^{\beta-1} \left\{ \frac{2g(b)}{\sigma\sqrt{\tau}} e^{-\frac{1}{2}d^2(1/x)} - \beta \int_{-\infty}^{d(1/x)} e^{-\frac{\xi^2}{2}} g(be^{[d(1/x)-\xi]\sigma\sqrt{\tau}}) d\xi \right\} +$$

$$+ \frac{e^{-q\tau}}{\sqrt{2\pi}} \left\{ \int_{-\infty}^{d_+(x)} e^{-\frac{\xi^2}{2}} g'(be^{[d_+(x)-\xi]\sigma\sqrt{\tau}}) d\xi + x^{\beta-2} \int_{-\infty}^{d_+(1/x)} e^{-\frac{\xi^2}{2}} g'(be^{[d_+(1/x)-\xi]\sigma\sqrt{\tau}}) d\xi \right\},$$

$$d_+(x) := d(x) + \sigma\sqrt{\tau} = (\sigma\sqrt{\tau})^{-1}[\ln x + (r-q-a+\sigma^2/2)\tau]. \tag{4.9}$$

**Theorem 4.3.** (*Lower and upper estimates of derivative*) *If $g(S)$ is continuous, piecewise differentiable, and $g(S)$ and $g'(S)$ are bounded, then we have the the following lower and upper estimates of the derivatives of the solutions $V$ of (4.1), (4.2), (4.3).*

$$F_1(S,t) \leq V_S(S,t) \leq F_2(S,t), \quad S > be^{-a(T-t)}, 0 \leq t < T \tag{4.10}$$

*Here*

$$F_1(S,t) = \frac{e^{-r\tau}}{be^{-a\tau}} x^{\beta-1} \left\{ \frac{2e^{-d^2(1/x)/2}}{\sigma\sqrt{2\pi\tau}} g(b) + N[d(1/x)]\min(-\beta \cdot g(S)) \right\}$$

$$+ e^{-q\tau} \{N[d_+(x)] + x^{\beta-2} N[d_+(1/x)]\} \min g'(S)$$

$$F_2(S,t) = \frac{e^{-r\tau}}{be^{-a\tau}} x^{\beta-1} \left\{ \frac{2e^{-d^2(1/x)/2}}{\sigma\sqrt{2\pi\tau}} g(b) + N[d(1/x)]\max(-\beta \cdot g(S)) \right\} \quad (4.11)$$
$$+ e^{-q\tau}\{N[d_+(x)] + x^{\beta-2} N[d_+(1/x)]\}\max g'(S)$$

From Theorems 3.2 to 3.6, we have the following strict monotonicity result.

**Theorem 4.4**. (*Strict monotonicity*) *Under the assumption of Theorem* 4.2, *assume that* $g(b) > 0$, $g'(S) \geq 0$. *Then for the solution* $V$ *to the problem* (4.1), (4.2), (4.3), *when* $\beta \leq 0$, *we have always* $V_S(S,t) > 0$ $(S > be^{-a(T-t)})$; *when* $\beta > 0$, *if* $g(b) \leq g(S) \leq g(+\infty) < +\infty$, $g'(S) \leq D < +\infty$ *and*

$$g(b) > \frac{\beta\sigma\sqrt{T}e^{\beta^2\sigma^2 T/8}}{2} \int_{-\infty}^{-\beta\cdot\sigma\sqrt{T}/2} e^{-\xi^2/2} g(be^{[-\beta\cdot\sigma\sqrt{T}/2-\xi]\sigma\sqrt{T}}) d\xi = \frac{\beta}{2}\int_0^\infty e^{-\frac{\eta^2}{2\sigma^2 T} - \frac{\beta\eta}{2}} g(be^\eta) d\eta \quad (4.12)$$

*then* $V_S(S,t) > 0$ $(S > be^{-a(T-t)})$ *holds. In particular, if*

$g(+\infty) \leq A(-\beta\sigma\sqrt{T}/2) \cdot g(b)$ (*where* $A(\xi)(>1)$ *is the function defined in Section 2*) (4.13)

*then* (4.12) *is true*.

## 5. Applications: Analysis on One Factor Structural Model for Puttable Bond with Credit Risk and Pricing Formulae

In this section, applying the results in the previous sections, we introduce one factor structural model for pricing puttable bond with credit risk [OKC-2021], prove that the bond price function strictly increases, and financially analyzes the price formula given in [OKC-2021] to derive representations of the survival probability, the early redemption probability of and the default probability.

### 5.1. One Factor Structural Model for Puttable Bond with Credit Risk

One-factor structural model for pricing corporate bond that enables early redemption on a specific day prior to maturity [OKC-2021] is as follows.

**Assumption 1**: Short-term interest rate $r$ is constant.

**Assumption 2**: The firm value $V$ consists of $m$ shares of stocks $S$ and $n$ sheets of zero coupon bonds:

$$V_t = mS_t + nB_t \quad (5.1)$$

and $V$ follows the geometric Brownian motion

$$dV = rVdt + \sigma VdW \quad \text{(under the risk neutral measure)}. \quad (5.2)$$

Here volatility $\sigma$ is a constant and $W$ is a Wiener process.

**Assumption 3**: Expected default occurs when

$$V \leq be^{-a(T-t)} \quad (a, b: \text{constants}).$$

**Assumption 4**: Default recovery is $R_d = R \cdot e^{-r(T-t)}$ ($0 \leq R < 1$:constant).

**Assumption 5**: At a particular time $T_1 (0 < T_1 < T)$ prior to maturity, the bond holder has the right to receive cash $E$ (unit of currency) and cancel the bond contract. This condition is called the *early redemption provision*.

**Assumption 6**: Our Corporate bond price is given by a sufficiently smooth deterministic function $B(V, t)$. The face value of the bond (the price at maturity $T$) is 1(unit of currency).

Such bonds are *puttable bonds* with credit risk.

### 5.2. Analytical pricing formula of puttable bond and strict increasing property

First, in the time interval $(T_1, T]$, the bond has no right of early redemption, so in this interval the bond is called the *straight bond*. In $(T_1, T]$, $B(V, t)$ is the solution of the following terminal boundary value problem of the Black-Scholes equation [OW-2013, OKC-2021].

$$\frac{\partial B}{\partial t} + \frac{1}{2}\sigma^2 V^2 \frac{\partial^2 B}{\partial V^2} + rV\frac{\partial B}{\partial V} - rB = 0, \ be^{-a(T-t)} < V < +\infty, \ T_1 < t < T \quad (5.3)$$

$$B(be^{-a(T-t)}, t) = e^{-r(T-t)}R, \ T_1 < t < T \quad (5.4)$$

$$B(V, T) = 1, \ V > V_b(T) = b. \quad (5.5)$$

Let $C(V, t) = B(V, t) - R \cdot e^{-r(T-t)}$, then $C(V, t)$ satisfies (5.3) and the following boundary conditions :

$$C(be^{-a(T-t)}, t) = 0, \ 0 < t < T \quad (5.6)$$

$$C(V, T) = 1 - R =: g(V), \ V > b. \quad (5.7)$$

$g(V) \equiv g(b)$ (constant), $g'(V) \equiv 0$ and from Theorem 4.1 and Theorem 4.2, we have (To simplify the notation below, we write $x = V/[be^{-a(T-t)}]$.)

$$C(V, t) = \frac{(1-R) \cdot e^{-r\tau}}{\sigma\sqrt{2\pi\tau}} \int_1^{+\infty} \left\{ e^{-\frac{1}{2}d^2(x/z)} - x^\beta e^{-\frac{1}{2}d^2(x^{-1}/z)} \right\} \frac{1}{z} dz =$$

$$= (1-R) \cdot e^{-r\tau} \{N(d(x)) - x^\beta N[d(1/x)]\} = (1-R) \cdot e^{-r\tau} \cdot W(V, \tau)$$

$$B(V, t) = R \cdot e^{-r\tau} + C(V, t) = e^{-r(T-t)}[R + (1-R) \cdot W(V, \tau)], T_1 < t < T. \quad (5.8)$$

Here

$$\tau = T - t, \ d(x, t) := \frac{\ln x + (r - a - \sigma^2/2)\tau}{\sigma\sqrt{\tau}}, \ \beta := 1 - 2(r-a)/\sigma^2.$$

$$B_V = C_V = \frac{e^{-r\tau}}{\sqrt{2\pi}be^{-a\tau}} x^{\beta-1} \left\{ \frac{2g(b)}{\sigma\sqrt{\tau}} e^{-d^2(1/x)/2} - \beta \cdot g(b) \int_{-\infty}^{d(1/x)} e^{-\xi^2/2} d\xi \right\}$$

$$= \frac{2(1-R)e^{-r\tau}}{be^{-a\tau}\sigma\sqrt{\tau}} x^{\beta-1} \left\{ \frac{1}{\sqrt{2\pi}} e^{-d^2(1/x)/2} - \frac{\beta\sigma\sqrt{\tau}}{2} N[d(1/x)] \right\}.$$

From Theorem 4.4, $C_V = B_V > 0 (V > be^{-a\tau}, T_1 \le t < T)$. And from Theorem 3.2 we have

$$0 < W(V, \tau) = N[d(x)] - x^\beta N[d(1/x)] < 1, \tag{5.9}$$
$$e^{-r\tau} R = B(be^{-q\tau}, t) < B(V, t) < B(+\infty, t) = e^{-r\tau}, \ V > be^{-a\tau}.$$

Next, since at time $T_1$ the bond holder considers early redemption, our bond price $B_0(V, t)$ in the time interval $[0, T_1]$ satisfies (5.3) and (5.4) in the domain $\{(V, t) \mid 0 < t < T_1, \ be^{-a(T-t)} < V < +\infty\}$, and the following terminal value condition.

$$B_0(V, T_1) = \max\{E, B(V, T_1)\}, \ V > be^{-a(T-T_1)}.$$

From (5.9), we have

$$e^{-r(T-T_1)} R = B(be^{-q(T-T_1)}, T_1) < B(V, T_1) < B(+\infty, T_1) = e^{-r(T-T_1)}, \ V > be^{-q(T-T_1)}$$

and since $B_V > 0$, if $e^{-r(T-T_1)} R < E < e^{-r(T-T_1)}$, then there exists a unique $K > be^{-a(T-T_1)}$ such that $B(K, T_1) = E$ and

$$B(V, T_1) < E \ \text{if} \ V < K; \ B(V, T_1) \geq E \ \text{if} \ V \geq K.$$

($K$ is called the *early redemption boundary* at time $T_1$.) Thus we can write as follows:

$$B_0(V, T_1) = \max\{E, B(V, T_1)\} = E \cdot 1\{V < K\} + B(V, T_1) \cdot 1\{V > K\}. \tag{5.10}$$

As in the above, let $C_0(V, t) = B_0(V, t) - R \cdot e^{-r(T-t)}$, then $C_0(V, t)$ satisfies (5.3) and (5.6) in the domain $\{(V, t) \mid 0 < t < T_1, be^{-a(T-t)} < V < +\infty\}$ and the boundary conditions:

$$C_0(V, T_1) = E \cdot 1\{V < K\} + B(V, T_1) \cdot 1\{V > K\} - R \cdot e^{-r(T-T_1)} := g_1(V), \ V > be^{-a(T-T_1)}. \tag{5.11}$$

From (5.8), we have

$$g_1(V) = [E - R \cdot e^{-r(T-T_1)}] \cdot 1\{V < K\} +$$
$$+ (1-R) \cdot e^{-r(T-T_1)} \left\{ N\left[d\left(\frac{V}{be^{-a(T-T_1)}}\right)\right] - \left(\frac{V}{be^{-a(T-T_1)}}\right)^{1-\frac{2(r-a)}{\sigma^2}} N\left[d\left(\frac{be^{-a(T-T_1)}}{V}\right)\right] \right\} \cdot 1\{V > K\}.$$

Directly calculating $C_0(V, t)(t < T_1)$ by Theorem 4.1 and using a method similar to [Bu-2004], we can find a representation of the price $B_0(V, t) = C_0(V, t) + R \cdot e^{-r(T-t)}$ of the puttable bond in the time interval $[0, T_1]$ (see [OKC-2021]). According to [OKC-2021], it is provided as follows:

$$B_0(V, t) = e^{-r(T-t)}(W - J) + E \cdot e^{r(T_1-t)} I + R \cdot e^{-r(T-t)}(1 - W + J - I), \ 0 \leq t < T_1 \tag{5.12}$$

Here

$$W(V, \ t) = N(d) - x^\beta N(\tilde{d}) \ (0 \leq t < T)$$
$$I(V, t) := N(b_1) - N(b_2) - x^\beta [N(\tilde{b}_1) - N(\tilde{b}_2)] \ (0 \leq t < T_1)$$
$$J(V, t) := N_2(d, b_1; \rho) - N_2(d, b_2; \rho) - x^\beta [N_2(\tilde{d}, \tilde{b}_1; \rho) - N_2(\tilde{d}, \tilde{b}_2; \rho)]$$
$$+ N_2(d, -b_1; -\rho) - N_2(d, -b_3; -\rho) - x^\beta [N_2(\tilde{d}, -\tilde{b}_1; -\rho) - N_2(\tilde{d}, -\tilde{b}_3; -\rho)]\} \ (0 \leq t < T_1)$$
$$d = \frac{\ln\frac{V}{be^{-a(T-t)}} + (r - a - \frac{\sigma^2}{2})(T-t)}{\sigma\sqrt{T-t}}, \ \tilde{d} = \frac{\ln\frac{be^{-a(T-t)}}{V} + (r - a - \frac{\sigma^2}{2})(T-t)}{\sigma\sqrt{T-t}} \ (t < T)$$

$$b_1 = \frac{\ln\frac{V}{be^{-a(T-t)}} + (r - a - \sigma^2/2)(T_1 - t)}{\sigma\sqrt{T_1 - t}}, \quad \tilde{b}_1 = \frac{\ln\frac{be^{-a(T-t)}}{V} + (r - a - \sigma^2/2)(T_1 - t)}{\sigma\sqrt{T_1 - t}}$$

$$b_2 = \frac{\ln\frac{V}{Ke^{a(T-T_1)}e^{-a(T-t)}} + (r - a - \frac{\sigma^2}{2})(T_1 - t)}{\sigma\sqrt{T_1 - t}}, \quad \tilde{b}_2 = \frac{\ln\frac{b^2 e^{-a(T-t)}}{Ke^{a(T-T_1)} \cdot V} + (r - a - \sigma^2/2)(T_1 - t)}{\sigma\sqrt{T_1 - t}}$$

$$b_3 = \frac{\ln\frac{Ke^{a(T-T_1)} \cdot V}{b^2 e^{-a(T-t)}} + (r - a - \frac{\sigma^2}{2})(T_1 - t)}{\sigma\sqrt{T_1 - t}}, \quad \tilde{b}_3 = \frac{\ln\frac{Ke^{a(T-T_1)}e^{-a(T-t)}}{V} + (r - a - \frac{\sigma^2}{2})(T_1 - t)}{\sigma\sqrt{T_1 - t}}$$

$$\rho = \sqrt{(T_1 - t)/(T - t)} \quad (t < T_1).$$

And since

$$g_1'(V) = C_V(V, T_1) \cdot 1\{V > K\} =$$

$$= \frac{2(1 - R)e^{-r(T-T_1)}}{be^{-a(T-T_1)}} \left(\frac{V}{be^{-a(T-T_1)}}\right)^{\beta - 1} \left\{ \frac{e^{-\frac{1}{2}d^2(be^{-a(T-T_1)}/V)}}{\sqrt{2\pi}\sigma\sqrt{T - T_1}} - \frac{\beta}{2} N\left[d\left(\frac{be^{-a(T-T_1)}}{V}\right)\right] \right\} \cdot 1\{V > K\} \geq 0,$$

$g_1(V)$ is a continuous, piecewise differentiable and increasing function, so we have

$$E - R \cdot e^{-r(T-T_1)} = g_1(be^{-a(T-T_1)}) \leq g_1(V) \leq g_1(+\infty) = e^{-r(T-T_1)}(1 - R).$$

We apply the condition (4.13) to the model (5.3), (5.6), (5.11) for $C_0(V, t)$. If $\beta \leq 0$, then we always have $\partial_V C_0(V, t) = \partial_V B_0(V, t) > 0 (t < T_1)$. Since

$$g_1(+\infty) \leq A(-\beta\sigma\sqrt{T}/2) \cdot g_1(be^{-a(T-T_1)}) \Leftrightarrow e^{-r(T-T_1)}(1 - R) \leq A(-\beta\sigma\sqrt{T}/2)(E - R \cdot e^{-r(T-T_1)}),$$

if $\beta > 0$, then $\partial_V C_0(V, t) = \partial_V B_0(V, t) > 0 (t < T_1)$ when

$$E \geq R \cdot e^{-r(T-T_1)} + (1 - R)e^{-r(T-T_1)} / A(-\beta\sigma\sqrt{T}/2). \tag{5.13}$$

(Note that $1 < A(-\xi) < \infty$ by Lemma 2.2.)

**Remark 5.1.** If we use the pricing formula of the puttable bond with one early redemption date and its strict increasing property with respect to the firm value, we can derive the pricing formula of the puttable bond with several early redemption dates prior to the maturity by the similar method of [OKC-2021]. To do this, solution representations of Black-Sholes equations with maturity payoff functions including power functions and multidimensional normal distribution functions similar to [OK-2013, OC-2019, OKC-2021].

### 5.3. Survival, early redemption and the default probabilities for puttable bond

The price (5.8) of the straight bond in the interval $(T_1, T]$ can be written as

$$B(V, t) = e^{-r(T-t)}[R + (1 - R) \cdot W(V, \tau)] = e^{-r(T-t)}W + R \cdot e^{-r(T-t)}(1 - W) \tag{5.14}$$

and $W(V, t)$ is the solution to the following problem [OKC-2021]:

$$\frac{\partial W}{\partial t} + \frac{1}{2}\sigma^2 V^2 \frac{\partial^2 W}{\partial V^2} + rV\frac{\partial W}{\partial V} = 0, \quad be^{-a(T-t)} < V < +\infty, \ 0 < t < T \quad (5.15)$$

$$W(be^{-a(T-t)}, t) = 0, \quad 0 \le t < T$$

$$W(V, T) = 1, \quad V > b.$$

From the estimate (4.8)(or (5.9)) we have $0 \le W \le 1$.

**Remark 5.2.** From (5.14), for $t \in (T_1, T]$, $W(V,t)$ can be regarded as the probability that default event will not occur at or after time $t$ under the condition that default event has not occur until $t$ (the *survival probability*), $1-W$ the probability that default will occur at or after time $t$ under the same condition (the *default* probability), and the price at time $t$ of the straight bond without early redemption provision is an expectation of the discounted value of the payoff of the bond in the case that default event will not occur at or after time $t$ and the discounted value of default recovery in the case that default event will occur at or after time $t$. (5.15) is the PDE model for survival probability for the straight bond when the default barrier is $be^{-a(T-t)}$.

Furthermore, using the estimate (4.8), we can also analyze the financial meaning of the pricing formula (5.12) of the puttable bond.

At $t = T_1$, we have $b_1 = b_3 = +\infty, \tilde{b}_1 = \tilde{b}_2 = -\infty$, and
$$V > K \Rightarrow b_2 = +\infty, \tilde{b}_3 = -\infty \quad \text{and} \quad V < K \Rightarrow b_2 = -\infty, \tilde{b}_3 = +\infty.$$

Thus

$$I(V, T_1) = 1\{be^{-a(T-T_1)} < V < K\},$$
$$J(V, T_1) = 1\{be^{-a(T-T_1)} < V < K\} \cdot W(V, T_1).$$

So $I(V,t)(t<T_1)$ is the solution to the problem (5.14) in domain $\{(V,t) \mid be^{-a(T-t)} < V < +\infty, \ 0 < t < T_1\}$ with terminal condition $I(V, T_1) = 1\{be^{-a(T-T_1)} < V < K\}$. From (4.8), we have $0 < I(V,t) < 1 \ (t < T_1)$. Similarly, $W - J \ (t < T_1)$ is the solution to the same problem with the terminal condition $(W-J)(V, T_1) = 1\{V > K\}W(V, T_1)$, and thus we have $0 < (W-J)(V,t) < 1 \ (t < T_1)$. And $1 - W + J - I \ (t < T_1)$ is the solution to the same problem with the terminal condition $(1-W+J-I)(V, T_1) = 1\{V > K\}[1-W(V, T_1)]$ and thus we have $0 < (1 - W + J - I)(V,t) < 1 \ (t < T_1)$.

**Remark 5.3.** Thus, in (5.12), for $t \in [0, T_1)$ $W - J$ can be regarded as the probability that default event will not occur at or after time $t$ and early redemption will not occur at time $T_1$ under the condition that default event has not occur until $t$ (it is called the *survival probability* of the puttable bond in $[0, T_1)$), $I$ the probability that default event will not occur at or after time $t$ but the bond holder will require early redemption at time $T_1$ under the same condition (it is simply called the *early redemption probability*), and $1 - W + J - I$ the probability that early redemption will not occur at time $T_1$ but default will occur at or after time $t$ under the same condition (the *default probability* of the puttable bond in $[0, T_1)$ ).

## 6. Conclusions

1) The problems on solution representations of boundary problems for the Black-Sholes equations with moving boundary of the form of exponential functions, min-max and gradient estimates reduce to a special fixed boundary value problem (3.1)~(3.3).

2) The strict monotonicity of the solutions to the special fixed boundary value problems for the Black-Sholes equation is not easily obtained even in the special case with constant maturity payoff functions, it is obtained using some strict estimates on cumulative density function of normal distribution and its derivative (Lemma 2.1~2.3). And in this case the convexity of solution is not always true (Theorem 3.2 and Remark 3.1). In the case with general increasing maturity payoff functions, we have the strict monotonicity of the solution when the difference between values on boundaries is not so large (Corollary 3.5 and Theorem 3.6).

3) In mathematical and financial analysis on the structural model for pricing corporate bonds with early redemption and its solutions, the above results of solution representation to boundary problems for the Black-Sholes equation, min-max estimates and strict monotonicity are used in essential.

**References**

[Agl-2016] Agliardi R., *Boundary value problems for PDEs arising in the valuation of structured financial products*, Pliska Studia Mathematica, 26, 2016, 83–98.

[BGW-1996] Bergman, Y.Z., Grundy B.D. and Wiener Z., *General Properties of Option Prices*, The Journal of Finance, **51**, 1996, 1573 – 1610, DOI:10.1111/j.1540-6261.1996.tb05218.x

[BS-1973] Black, F. and Scholes, M., *The pricing of options and corporate liabilities*. Journal of Political Economics, **81**, 1973, 637-654.

[Bu-2004] Buchen, P., *The Pricing of dual-expiry exotics*, Quantitative Finance, 4 (2004), 101-108.

[Bu-2001] Buchen, P., *Image Option and the Road to Barries*, Risk , 14:9(2001) , 127~130

[CR-1976] Cox, J.C. and Ross S.A., *A Survey of Some New Result in Financial Option Pricing Theory*, The Journal of Finance, **31**, 1976, 383 – 402, DOI: 10.1111/j.1540-6261.1976.tb01893.x

[FNV-2016] Falcó, A., Ll. Navarro, C. Vázquez, *A Direct LU Solver for Pricing American Bond Options under Hull-White model*, to appear in Journal of Computational and Applied Mathematics (2016), http://dx.doi.org/10.1016/j.cam.2016.05.003

[Mer-1973] Merton, R.C., *Theory of rational option pricing*, The Bell Journal of Economics and Management Science, 4, 1973, 141-183. Cross-ref

[Jia-2005] Jiang, Li-shang, Mathematical Models and Methods of Option Pricing, World Scientific, 2005.

[OW-2013] O, H.C., Wan, N., *Analytical Pricing of Defaultable Bond with Stochastic Default Intensity*, arxiv:1303.1298v3[q-fin.PR], (2013), 1-35.

[OJK-2016] O, H. C., J.J.Jo, J.S.Kim, *General properties of solutions to inhomogeneous Black–Scholes equations with discontinuous maturity Payoffs*, J. Differential Equations 260 (2016), 3151–3172.

[OJKJ-2017] O, H-C.; Jo, J.J.; Kim, S.Y.; Jon, S.G.; *A comprehensive unified model of structural and reduced form type for defaultable fixed income bonds*, Bulletin of the Iranian Mathematical Society , 43, 3 (2017) 575-599.

[OC-2019] O, H.C., D.S.Choe, *Pricing Formulae of Power Binary and Normal Distribution Standard Options and Applications*, arXiv:1903.04106v1[q-fin.PR] 11 Mar (2019), 1-24.

[OK-2013] O, H.C., M.C. Kim, *Higher order binary options and multiple-expiry exotics*, Electronic Journal of Mathematical Analysis and Applications, Vol. 1(2) July (2013), pp. 247-259.

[OKC-2021] O, H.C., T.S. Kim, T.S. Choe, *Solution Representations of Solving Problems for the Black-Scholes equations and Application to the Pricing Options on Bond with Credit Risk*, arXiv:2109.10818v1 [q-fin.PR] 17 Aug (2021), 1-12..

[OKP-2014] O, H.C., Kim, D.H., Pak, C.H., *Analytical pricing of defaultable discrete coupon bonds in unified two-factor model of structural and reduced form models*, Journal of Mathematical Analysis and Applications, Vol. **416**, 2014, 314 – 334.

[RF-2016] Russo V, Fabozzi FJ. *Pricing coupon bond options and swaptions under the one- factor Hull-White model*. J Fixed Income, 25 (2016) 76–82

[RF-2017] Russo V, Fabozzi F., *Pricing coupon bond options and swaptions under the two- factor Hull–White model*. J Fixed Income 27 (2017) 30–36

[ZW-2012] Zhang, K., S. Wang, *Pricing American bond options using a penalty method*, Automatica, 48 (2012), 472–479.

[ZYL-2011] Zhou, H. J., K. F. C. Yiu, L. K. Li, *Evaluating American put options on zero

- *coupon bonds by a penalty method*, Journal of Computational and Applied Mathematics, 235 (2011), 3921–3931.